\documentclass[11pt,letterpaper]{article}
\pdfoutput=1

\usepackage{graphicx,array}
\usepackage[dvipsnames]{xcolor}
\definecolor{darkblue}{rgb}{0,0.1,0.5}
\definecolor{darkgreen}{rgb}{0,0.5,0.2}
\definecolor{darkred}{RGB}{153,26,0}
\usepackage{ulem}
\usepackage{slashed}
\definecolor{seablue}{rgb}{0,0.2,0.6}
\usepackage{latexsym}
\usepackage{amssymb, amsmath}
\usepackage{slashed}
\usepackage{bm}        
\usepackage[numbers,sort&compress]{natbib}
\usepackage{bm,relsize}
\usepackage{slashed}
\definecolor{viola}{RGB}{134,41,198}
\usepackage{mathrsfs}
\usepackage{hyperref} 
\hypersetup{
    colorlinks=true,       
    linkcolor=darkblue,          
    citecolor=BrickRed,        
    filecolor=magenta,      
    urlcolor=MidnightBlue           
}
\usepackage[all]{hypcap} 
\usepackage{subcaption}

\usepackage{listings}
\usepackage{xcolor}

\definecolor{codebg}{rgb}{0.95,0.95,0.95}

\usepackage{colortbl}

\usepackage[font=small,labelfont=bf,labelsep=period]{caption}

\newcommand{\be}{\begin{equation}}
\newcommand{\ee}{\end{equation}}
\newcommand{\Hc}{\mathcal{H}}
\newcommand{\Mpl}{M_{\rm Pl}}
\newcommand{\llike}{-\log\mathscr{L}_{\rm min}}

\usepackage{tikzfeynman}

\begin{document}

\begin{flushright}

\end{flushright}
\vspace{.6cm}
\begin{center}
{\LARGE \bf 
Consistent Theories for the DESI dark energy fit
}\\
\bigskip\vspace{.7cm}
{
\large Alessio Notari$^{a,b}$, Michele Redi$^c$, Andrea Tesi$^c$}
\\[7mm]
 {\it \small
$^a$ Departament de F\'isica Qu\`antica i Astrofis\'ica \& Institut de C\`iencies del Cosmos (ICCUB), \\
Universitat de Barcelona, Mart\'i i Franqu\`es 1, 08028 Barcelona, Spain\\
\vspace{.1cm}
$^b$ Galileo Galilei Institute for theoretical physics, Centro Nazionale INFN di Studi Avanzati\\
Largo Enrico Fermi 2, I-50125, Firenze, Italy\\ 
\vspace{.1cm}
$^c$INFN Sezione di Firenze, Via G. Sansone 1, I-50019 Sesto Fiorentino, Italy\\
Department of Physics and Astronomy, University of Florence, Italy
 }
\end{center}

\vspace{.2cm}

\centerline{\bf Abstract} 
\begin{quote}
We search for physically consistent realizations of evolving dark energy suggested by the cosmological fit of DESI, Planck and Supernovae data.
First we note that any lagrangian description of the standard Chevallier-Polarski-Linder (CPL) parametrization for the dark energy equation of state $w$, allows for the addition of a cosmological constant.  We perform the cosmological fit finding new regions of parameter space that however continue to favour dark energy with $w<-1$ at early times, that is challenging to realize in consistent theories. 
Next, in the spirit of effective field theories, we consider the effect of higher order terms in the Taylor expansion of the equation of state of dark energy around the present epoch. 
We find  that non-linear corrections of the equation of state are weakly constrained, thus opening the way to scenarios that differ from CPL at early times, possibly with $w>-1$ at all times. We present indeed scenarios where evolving dark energy can be realized through quintessence models. We introduce in particular the ramp model where dark energy coincides with 
CPL at late times and approximates to a cosmological constant at early times. The latter model provides a much better fit than $\Lambda$CDM, and only slightly worse than $w_0w_a$CDM, but with the notable advantage of being described by a simple and theoretically consistent lagrangian of a canonical quintessence model.
\end{quote}

\vfill
\noindent\line(1,0){188}
{\scriptsize{ \\ E-mail:\texttt{   \href{notari@fqa.ub.edu}{notari@fqa.ub.edu}, \href{mailto:michele.redi@fi.infn.it}{michele.redi@fi.infn.it}, \href{andrea.tesi@fi.infn.it}{andrea.tesi@fi.infn.it}}}}
\newpage
\tableofcontents

\section{Introduction}
Like it or not, the large scale behavior of our Universe today is determined to a considerable extent by the role played by the Cosmological Constant (CC): just a number, its present abundance $\Omega_{\Lambda}$.  

It is fair to say, although with an important caveat, that a different CC, approximately, is also responsible for determining the initial stages of cosmic history during inflation. It is certainly ironic to see that Einstein's ``biggest blunder'' turned out to be instrumental in understanding the Universe at large scales, at late but also at early times.  However, the inflationary epoch and the present one differ substantially in the details. 

Indeed, the fact that inflation has to end implies that the constant is not constant in time after all. Such modulation, albeit rather slow, when interpreted in the framework of quantum field theory inevitably determines the existence of at least one new quantum degree of freedom: the inflaton fluctuation, in this case. This is a theoretical consideration that turned out to be extremely powerful and predictive for the initial conditions of our Universe. 

In the late universe, on the contrary, there are no theoretical considerations that require a modulation in time of the CC, therefore  we have to rely on \textit{data}. So far, no convincing evidence for a time variation of the CC has been found. The question, however, still stands as it would imply at the level of quantum physics the presence of at least one new degree of freedom.

Since the question is very fundamental, there are many cosmological surveys aimed at inspecting possible time dependencies of the CC. In this context, it is rather common to talk about dark energy (DE), as a proxy for time dependent effects in the CC. DE is a new component that might have a non-trivial time evolution and therefore associated new degrees of freedom. Since DE is expected to provide a small contribution to clustering - which is strongly constrained - it mainly can be detected by studying how it modifies the background evolution of the universe, through the expansion history. Also, since the CC-domination happens at redshift $z$ around 0.3, its effects are usually larger at low-$z$ where the determination of the Hubble parameter can be achieved thanks to Baryon-Acoustic-Oscillation (BAO) and Supernovae (SN) measurements.

Prompted by the first year data release of DESI \cite{DESI:2024mwx}, we explore under which circumstances the new data show evidence for a time-dependent `constant' (see also~\cite{Yin:2024hba, Yang:2024kdo, Wang:2024hks, Wang:2024rjd}). By combining different datasets, we explore several DE realizations defined by their equation of state and sound speed, that is important to assess the size of DE perturbations. A very useful parametrization is the CPL equation of state~\cite{Chevallier:2000qy,Linder:2002et}, that at face value has a better fit to data than $\Lambda$CDM model \cite{DESI:2024mwx}. Such a parametrization for the equation of state  is a linear expansion in the scale factor around the present epoch. Using CPL the data seems to prefer - by quite some margin - a model that `crosses the phantom divide', i.e. has an equation of state that goes below $-1$ at early times. An equation of state smaller than $-1$ is difficult to reconcile with the principles of quantum field theory, as it is often associated with the presence of ghosts or at least gradient instabilities.

In this work we challenge the idea that the CPL parametrisation has to be taken as a fully fledged model and instead use it as an effective description of the behavior of DE close to the redshifts where it is tested. As such, it may be possible to find models without pathologies that correctly match to CPL at the corresponding redshifts. 
We study in particular deformations of CPL including a CC that can be added to the action of any fluid as well as higher order terms in the equation of state.
We then consider explicit quintessence scenarios where $w>-1$.  Most importantly these scenario allow to include consistently perturbations that present no pathologies. 

In section \ref{sec:data} we present the cosmological observables used in our analysis and we discuss at length the dataset used in order to reproduce the results of \cite{DESI:2024mwx}. We perform a fit to the CPL model, and we discuss the impact of different SN datasets with a focus on Pantheon+ \cite{Pantheon+} and DES-SNYR5 \cite{DES:2024tys}, also constructing the likelihood for the latter, and the impact of DE perturbations. In section \ref{sec:cost} we discuss the physical implications of the CPL fit, first we add the CC to the model and then we show that present data have not enough constraining power to determine whether CPL is valid also at higher redshift.  This result is achieved by testing an effective expansion of the equation of state up to quadratic order in the scale factor. By taking inspiration from this, we   consider more generic deviations that however never cross the phantom divide. These are addressed in section \ref{sec:models} where we focus on models. We both consider k-essence and q-essence models. Our main result is that we find quintessence models that have good fit to data (almost as good as CPL) and are healthy quantum field theories. We also compute the corresponding potential, by reverse engineering the equation of state that gives the best fit to data. We conclude in section \ref{sec:conclusions}, leaving technical material for the appendix \ref{app:inputs}.

\begin{figure}[t]
  \centering
  \includegraphics[width=0.97\textwidth]{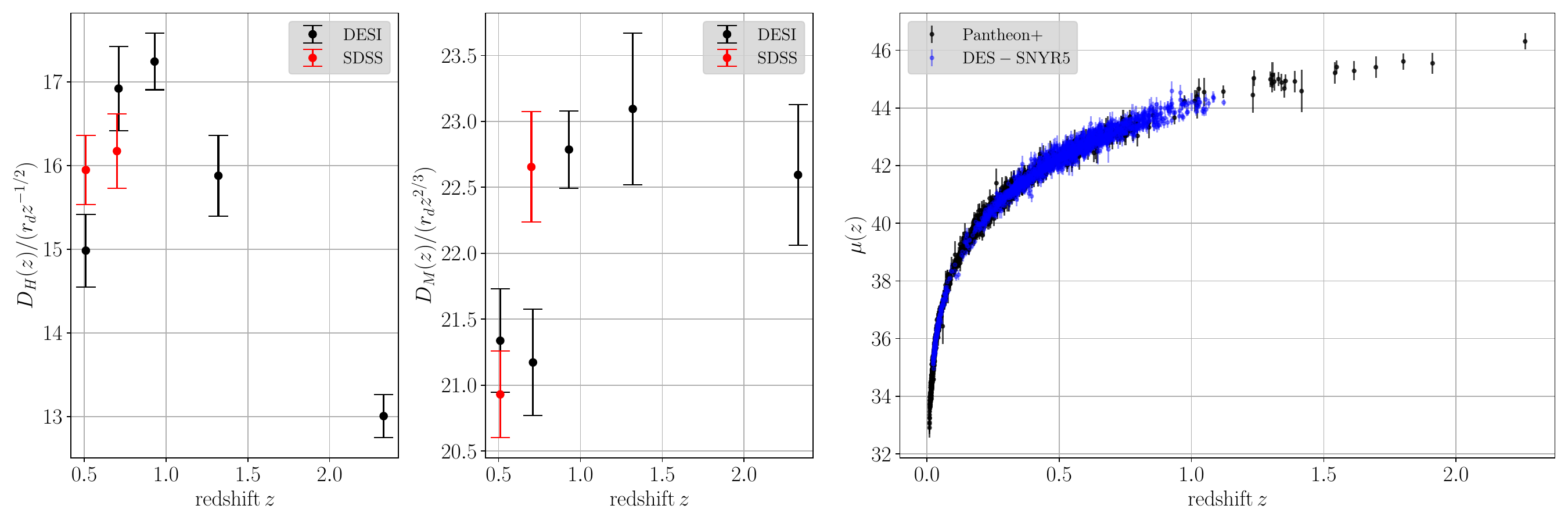}~
      \caption{\it~\textbf{ Cosmic ladders used in this work.} We include DESI-BAO and SN datasets. Here we also report the comparison between the BAO measurements of SDSS and DESI  (left and center panels), and the comparison between Pantheon+ and DES-SNYR5 (right panel). The nuisance parameter that determines the overall logarithmic normalization $\mu$ of supernovae has been set to its best fit value derived for the model of section \ref{sec:cost} for both Pantheon+ and DES-SNYR5 catalogues.}
  \label{fig:distances}
\end{figure}

\section{The late Universe as seen through distances}\label{sec:data}

We start reviewing the main observables that are useful to determine and constrain the late cosmological evolution of our universe. 

The quantities that are easy to compute theoretically are distances derived from the Hubble parameter $H(a)$ as a function of the scale factor $a=1/(1+z)$. In order to set the conventions, we express Hubble as
\be
H(a)^2=H_0^2\sum_i \Omega_i f_i(a)
\ee
where $f_i(a)$ are the solutions of the continuity equation for each given species contributing to the total energy budget,
\be\label{eq:f}
\frac {d\rho_i(a)}{da} = - \frac 3 a(1+w_i(a))\rho_i(a), \quad \quad f_i(a) \equiv \frac{\rho_i(a)}{\Omega_i \rho_{c,0}}\,.
\ee
where $\rho_{c,0}$ is the critical density today. 
We restrict our analysis to a spatially flat universe, so that the energy budget of all the species determines the size of the CC or of the dark-energy component under consideration as the remainder in $\sum \Omega_i=1$. 
These quantities  are background quantities, but also fluctuations of these variables are important when considering CMB and matter power spectrum data. We discuss them later.

There are many background observables that are sensitive to the ratio $H(z)/H_0$, and therefore to the background composition of the Universe. Such observables are all related to some extent to the \textit{transverse comoving distance} $D_M(z)$, which - for a spatially flat Universe - is computed to be
\be
D_M(z)\equiv c \int_0^z dz' \frac{1}{H(z')}\,.
\ee
Another important distance is the Hubble diameter $D_H$ related previous distance by $D_H(z)=\partial D_M(z)/\partial z$. These distances are redshift dependent and as such can be strongly constrained by means of observables sensitive to different redshift bins, such as CMB, BAO and Supernovae. 

Since BAO and CMB are sensitive to dimensionless angles in the sky, one needs to have as a reference a given `ruler'. For the CMB and BAO measurements, the rulers are the sound horizon at recombination $r_*\equiv r_s(z_*)$ and at baryon drag $r_d\equiv r_s(z_d)$. In general the sound horizon at a reference redshift $z$ is given by
\be
r_s(z)\equiv \int_{z}^\infty dz' \frac{c_s(z')}{H(z')} \, .
\ee
The angle that corresponds to acoustic peaks is $\theta=r(z_*)/D_M(z_*)$, which can be effectively measured in the CMB. Similarly BAO measure the ratio $r_d/D_M(z)$ at much lower redshifts. In other contexts, as in the case of Supernovae, calibration require extra ingredients. The flux from SN is given by $F=L/(4\pi d_L^2)$, where $d_L$ is the luminosity distance, which in any metric theory~\cite{Bassett:2003vu,EUCLID:2020syl}
is related to $D_M$ via $d_L(z)\equiv(1+z)D_M(z)$. The flux is often traded for a measurement of the magnitude $\mu(z)$, which s given by
\be
\mu(z)\equiv 5 \log_{10}[(1+z)D_M(z)] - M \, ,
\ee
where $M$ can be known only by setting the overall calibration: this is done in measurements that use the so-called distance ladder method~\cite{Riess:2021jrx} to determine $H_0$ from supernovae; however, here we do not rely on such measurements, which are linked to the well-known Hubble tension~\cite{DiValentino:2020zio} and have been addressed in the context of DESI fits in~\cite{Allali:2024cji}, and we marginalize over $M$. We will return to this in section~\ref{sec:models}.

Thus the following set of observables  emerges, 
\be
\mathrm{BAO}:\quad \frac{D_M(z)}{r_d},\quad \frac{D_H(z)}{r_d}\,;\quad\quad\quad \mathrm{SN}: \quad \frac{D_M(z)}{10^{M/5}}\, ,
\ee
and BAO+SN can therefore constrain the background composition of the Universe if $r_d$ is known (by combining with CMB) and provided enough statistics is obtained for several different redshifts. Given the different dependence on $D_M(z)$, BAO and SN offer complementary information. The dataset collected are shown in figure \ref{fig:distances} for  BAO and Supernovae.

\subsection{Evolving dark energy: CPL parametrization}
Combinations of DESI BAO data with cosmological datasets that include CMB+ SN disfavour the standard  $\Lambda$CDM cosmological model when the data are fitted to a model with non-constant dark energy \cite{DESI:2024mwx}. 

The reference model for this kind of late time modification of the Universe is the so-called CPL parametrization of dark energy \cite{Chevallier:2000qy,Linder:2002et}, where in addition to Cold Dark Matter (CDM) the cosmological model is supplemented by a DE component. In the original  CPL parametrization the CC is set to zero $\Omega_{\Lambda}=0$, such that the closure of the energy budget is guaranteed by $\Omega_{\rm CPL}=1-\sum_i \Omega_i$, where the sum runs on all the SM species and CDM. The CPL component is defined by the following time-dependent equation of state,
\be\label{eq:CPL}
w_{\rm CPL}(a)=w_0 + w_a(1-a)\,.
\ee
With this expression is possible to solve eq.~\eqref{eq:f}, which provides the following (normalized) dark energy density
\be\label{eq:fCPL}
f_{\rm CPL}= \exp{(3 w_a (a-1))} a^{-3(1+w_0+w_a)}\,.
\ee
One of the virtues of the CPL parametrization is that for $w_0+w_a\le 0$ the energy density becomes negligible at early times.
This expression is the one needed to compute $H(z)$ at all times, and in turn to determine the cosmological distances introduced above. At the level of background cosmology, this is all the effect of the new CPL component. 

In order to have a complete description one should also include fluctuations of the  CPL component. 
These are in general described by the following equations for its density contrast $\delta$ and its velocity divergence $\theta$ in conformal newtonian gauge \cite{MB}
\begin{eqnarray}\label{perfect}
\delta' + (1+w)(\theta -3 \Phi')+3 \Hc \delta \left(c_s^2-w\right ) &=&0\, ,\\
\theta' + \Hc (1-3w)\theta +\frac{w'}{1+w} \theta -\frac{c_s^2}{1+w} k^2\delta &=& k^2\Psi\, \, .
\label{eq:perfectfluid}
\end{eqnarray}
with $\Phi$ and $\Psi$ the gravitational potentials and we neglected the shear-stress.

This system of equations is closed once the sound-speed $c_s^2\equiv \frac{\delta P}{\delta \rho}$ is given or computed.
We expect that, since for $w_0+w_a< 0$ this energy component is always negligible in the early Universe, these fluctuations might have a limited impact on the bulk of CMB data, but they potentially affect the integrated Sachs-Wolfe effect and CMB lensing (as well as the matter power spectrum). These equations become problematic
when $w$ crosses $-1$. For the CPL parametrization this corresponds to $w_0+w_a< -1$, which is actually the region favoured by the cosmological fit \cite{DESI:2024kob}.
To overcome this difficulty the \texttt{CLASS} code uses the prescription~\cite{Fang:2008sn} together with $c_s^2=1$, which is based on multiple scalar field models designed to cross the phantom divide. 

\bigskip
The results presented in the recent work by the DESI collaboration \cite{DESI:2024mwx} disfavor the $\Lambda$CDM point, $w_0=-1$, $w_a=0$, when allowing for the CPL fluid. In particular, when combining with CMB and SN, the $\Lambda$CDM point is more than three sigma away from the best-fit. To validate our procedure we first reproduced the results of \cite{DESI:2024mwx} as reported in appendix \ref{appA}. Next we  perform a bayesian inference of the cosmological parameters in various extensions of $\Lambda$CDM.

\section{Implications  for dark energy}\label{sec:cost}

In this section we assume that the DESI in combination with CMB and supernovae measurements indicate the existence of evolving dark energy
and try to interpret it within consistent quantum field theories. By consistent here we mean theories that
do not feature states with negative kinetic terms, i.e. ghosts. 

\bigskip
One approach is to take the CPL component at  face value. If so, we must conclude that dark energy crosses the ``phantom divide'', $w<-1$ at early times.
This is in general problematic because $w<-1$ typically implies the presence of ghosts. As emphasized in \cite{Creminelli:2008wc}
this conclusion could be avoided if sufficiently general theories are considered. In particular the phantom divide can be crossed without immediate inconsistencies 
in scalar theories with higher derivative terms, known as k-essence. Beside the question of  whether these theories admit a UV completion, their phenomenological viability is not obvious as we will review in section \ref{sec:models}. 

We thus explore other possibilities to reproduce the DESI and supernovae measurements.

\subsection{CPL + $\Lambda$}

\begin{figure}[t!]
\begin{center}
\begin{minipage}{.9\textwidth}
\centering
\includegraphics[width=.7\textwidth]{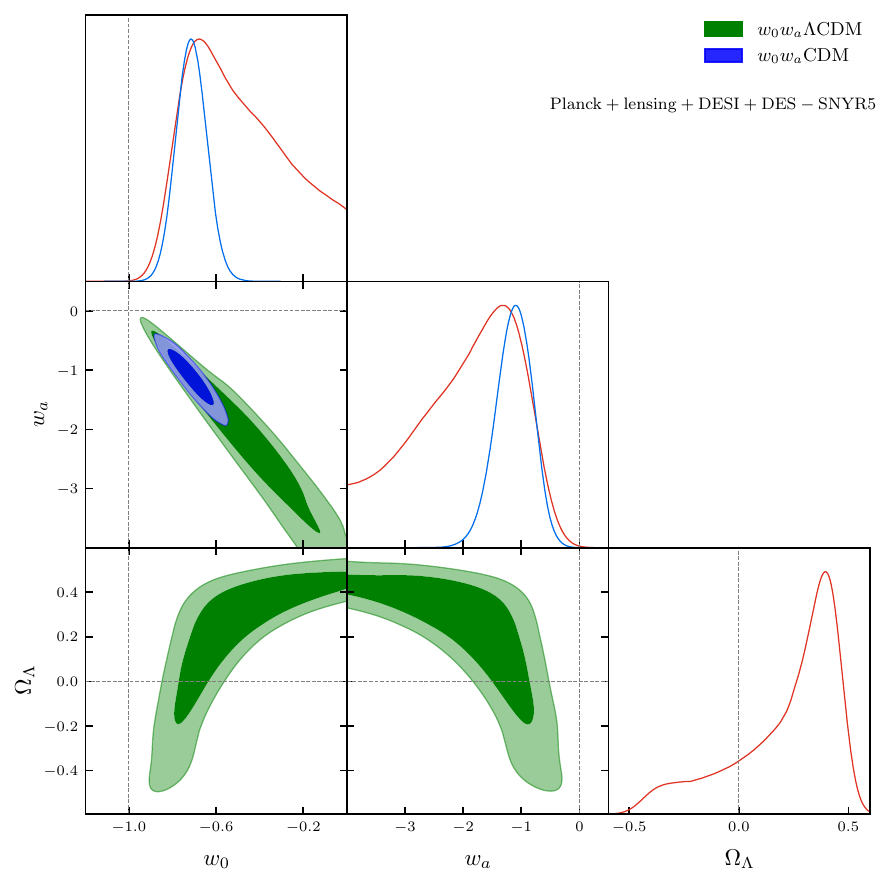}~
\end{minipage}

\begin{minipage}{\textwidth}
    \centering
    \renewcommand{\arraystretch}{1.1} 
    \begin{tabular}{l|c|c|c|c}
        \hline
        \textbf{Parameter} & \textbf{Best-fit} & \textbf{Mean $\pm \sigma$} & \textbf{95\% lower} & \textbf{95\% upper} \\ \hline
       
$100\omega_{\rm b }$ &$2.242$ & $2.238_{-0.014}^{+0.014}$ & $2.211$ & $2.266$ \\ 
$\omega_{\rm cdm }$ &$0.1193$ & $0.1196_{-0.00098}^{+0.00098}$ & $0.1177$ & $0.1216$ \\ 
$100\theta{}_{s }$ &$1.042$ & $1.042_{-0.00028}^{+0.00028}$ & $1.041$ & $1.042$ \\ 
$\log(10^{10}A_{s })$ &$3.045$ & $3.044_{-0.015}^{+0.014}$ & $3.015$ & $3.073$ \\ 
$n_{s }$ &$0.9685$ & $0.9654_{-0.0038}^{+0.0039}$ & $0.9579$ & $0.9729$ \\ 
$\tau_{\rm reio }$ &$0.0547$ & $0.05442_{-0.0076}^{+0.0071}$ & $0.03961$ & $0.06924$ \\ 
\hline
$w_0$ &$-0.7619$ & $-0.4342_{-0.41}^{+0.16}$ & $-0.8983$ & $0.1456$ \\ 
$w_a$ &$-0.9223$ & $-2.359_{-0.61}^{+1.8}$ &--& -- \\ 
$\Omega_{\Lambda }$ &$-0.1528$ & $0.237_{-0.054}^{+0.26}$ &--& -- \\ 
\hline\hline
$\Omega_{\rm CPL}$ &$0.8415$ & $0.448_{-0.28}^{+0.057}$ & -- & -- \\ 
$H_0$ [km/s/Mpc] &$67.64$ & $67.32_{-0.65}^{+0.62}$ & $66.04$ & $68.54$ \\ 
$M$ [DES nuisance]&$-0.03474$ & $-0.04259_{-0.016}^{+0.016}$ & $-0.07444$ & $-0.01099$ \\ 
 \hline\hline
                \multicolumn{5}{c}{$\llike = 2215$} \\ \hline
    \end{tabular}
\end{minipage}
 \caption{\it~\textbf{$\mathbf{w_0 w_a}\Lambda$CDM model with Planck18 TTTEEE+lensing+DESI+DES-SNYR5 datasets.} Marginalized posterior distributions for $(w_0,w_a,\Omega_\Lambda)$. We assume a prior $\Omega_{\Lambda}\in [-0.5,0.5]$, and uniform priors on the rest of the input parameters. We also report the bounds of the $95\%$ posterior interval.}
    \label{fig:LCPL}
    \end{center}
\end{figure}

We note that in any attempt to describe evolving dark energy within a lagrangian description a CC 
can always be added to the action. This implies that the dark energy density assuming a CPL equation of state evolves in general as
\begin{equation}
\Omega_{\rm DE} f_{\rm DE}(a)= \Omega_{\Lambda}+ \Omega_{\rm CPL} f_{\rm CPL}(a)\,,
\label{eq:LCPL}
\end{equation}
where we can choose $f_{\rm DE}(1)=1$ and $\Omega_{\rm DE}=\Omega_{\Lambda}+ \Omega_{\rm CPL}$ assuming spatial flatness.
The parameter $\Omega_{\Lambda}$ cannot be generically re-absorbed into the parameters $w_0$ and $w_a$ of the CPL parametrization
and therefore we see no reasons not to include the CC in the fit.

We thus perform a fit including the CPL fluid and CC, see \cite{Wang:2024hwd} for a similar study. Because the  CC does not carry degrees of freedom
the system can be equivalently described with a single fluid with a modified equation of state in order to reproduce (\ref{eq:LCPL}).
This is \cite{Adil:2023ara},
\begin{equation}
w_{\rm \Lambda CPL}\equiv -1 -\frac 1 3 \frac {d\log f_{\rm DE}}{d\log a} =\frac{e^{3 (a-1) w_a} (\Omega_{\rm CPL}-\Omega_\Lambda ) (-a w_a+w_0+w_a)-\Omega_\Lambda  a^{3
 (w_0+w_a+1)}}{\Omega_\Lambda  a^{3 (w_0+w_a+1)}+e^{3 (a-1) w_a} (\Omega_{\rm CPL}-  \Omega_\Lambda )} \, .
   \label{eq:LCPLeff}
 \end{equation}
We have studied this model including the cosmological constant in the CPL routine of CLASS, see Fig. \ref{fig:LCPL} for the posterior distributions.
In principle the addition of $\Omega_\Lambda$ could change the result of the fit and lead to a healthy CPL fluid with $w>-1$.
This is not what happens however, the posterior distribution continues to favour $w_0+w_a<-1$ crossing the phantom divide in the past. 
Moreover the minimum $\chi^2$ does not improve significantly. 

We note however that $\Omega_{\Lambda}$ is degenerate with other parameters in the fit. In fact the posterior distribution shows that
within errors $\Omega_\Lambda$ can be sizable and of with either signs. In fact the best fit is obtained  for a negative value of $\Omega_\Lambda$.
This can be understood as follows. Expanding  (\ref{eq:LCPLeff}) around $a=1$,
\begin{equation}
w_{\Lambda {\rm CPL}}= \frac{w_0(\Omega_{\rm CPL}-\Omega_\Lambda)- \Omega_\Lambda }{\Omega_{\rm CPL}}+\frac{(\Omega_{\rm CPL}-\Omega_\Lambda ) \left(3 (w_0+1)^2 \Omega_\Lambda +w_a \Omega_{\rm CPL}\right)}{\Omega_{\rm CPL}^2}(1-a)+\dots
\end{equation}
Therefore around $a=1$ the addition of $\Omega_\Lambda$ amounts to a redefinition of the parameters $w_0$ and $w_a$ in the CPL parametrization.
The degeneracy is thus associated to the fact that the fit is mostly sensitive to $w$ and $w'$ close to the present epoch, while CC modifies genuinely only higher derivative terms.
Note that the instantaneous equation of state is actually very precisely constrained (depending on the dataset) around some pivot scale at $0.25 \lesssim z \lesssim 0.3$ (see, e.g., Sec 5.2 of~\cite{DESI:2024mwx} and references therein and \cite{Cortes:2024lgw}, which discussed reparameterizations in terms of different pivot redshifts ).

This motivates the study of higher order terms in the equation of state to which we now turn.

\subsection{CPL as an effective theory}\label{sec:effectiveCPL}

\begin{figure}
\begin{center}
\begin{minipage}{.9\textwidth}
\centering
\includegraphics[width=.8\textwidth]{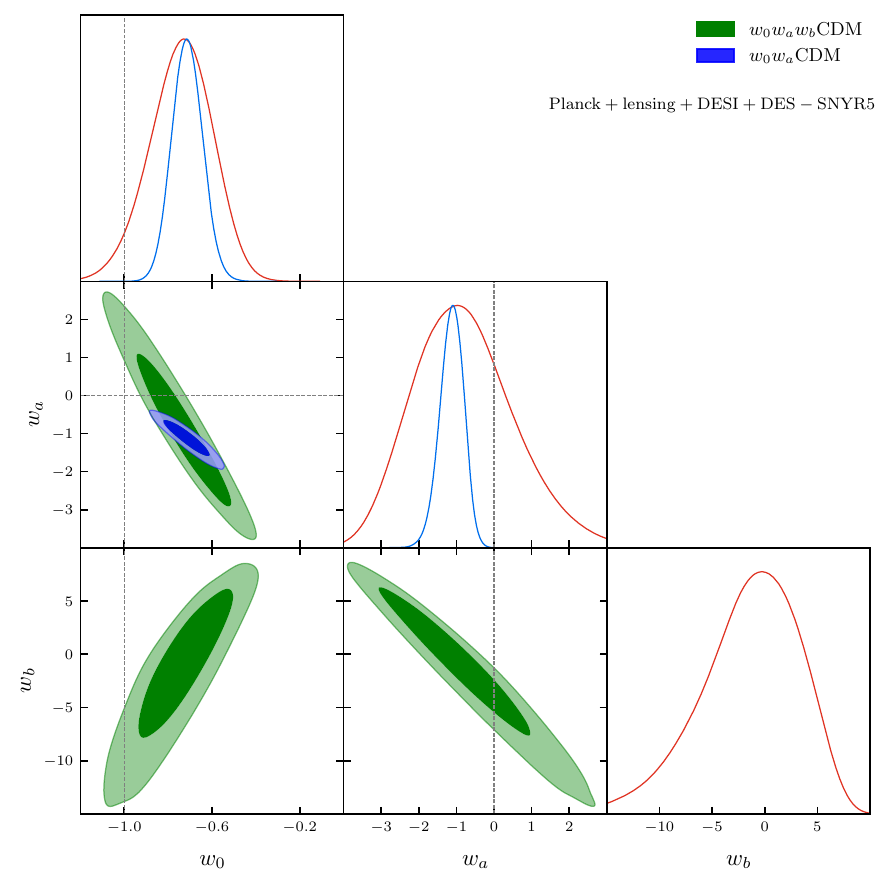}~
\end{minipage}

\begin{minipage}{\textwidth}
    \centering
    \renewcommand{\arraystretch}{1.1} 
    \begin{tabular}{l|c|c|c|c}
        \hline
        \textbf{Parameter} & \textbf{Best-fit} & \textbf{Mean $\pm \sigma$} & \textbf{95\% lower} & \textbf{95\% upper} \\ \hline
$100~\omega{}_{b }$ &$2.247$ & $2.236_{-0.014}^{+0.014}$ & $2.208$ & $2.263$ \\ 
$\omega{}_{\rm cdm }$ &$0.1201$ & $0.1199_{-0.001}^{+0.0011}$ & $0.1179$ & $0.122$ \\ 
$100*\theta{}_{s }$ &$1.042$ & $1.042_{-0.00029}^{+0.00029}$ & $1.041$ & $1.042$ \\ 
$\log(10^{10}A_{s })$ &$3.053$ & $3.042_{-0.015}^{+0.014}$ & $3.013$ & $3.071$ \\ 
$n_{s }$ &$0.9646$ & $0.9648_{-0.0039}^{+0.0039}$ & $0.9571$ & $0.9724$ \\ 
$\tau{}_{\rm reio }$ &$0.05591$ & $0.05321_{-0.0076}^{+0.0072}$ & $0.03868$ & $0.0683$ \\ 
\hline
$w_0$ &$-0.6863$ & $-0.7425_{-0.14}^{+0.16}$ & $-1.032$ & $-0.4638$ \\ 
$w_a$ &$-1.267$ & $-0.7822_{-1.6}^{+1.2}$ & $-3.387$ & $1.972$ \\ 
$w_b$ &$0.1298$ & $-1.465_{-3.7}^{+5.8}$ & $-11.28$ & $7.435$ \\ 
\hline\hline
$\Omega_{\rm CPL}$ &$0.6842$ & $0.6853_{-0.0078}^{+0.0084}$ & $0.6691$ & $0.702$ \\ 
$H_0$[km/s/Mpc] &$67.36$ & $67.42_{-0.76}^{+0.78}$ & $65.9$ & $68.98$ \\ 
        $M$ [DES nuisance]  &$-0.04017$ & $-0.03962_{-0.017}^{+0.017}$ & $-0.07272$ & $-0.005952$ \\ 
 \hline\hline
                \multicolumn{5}{c}{$\llike = 2215$} \\ \hline
    \end{tabular}
\end{minipage}
 \caption{\it~ \textbf{$\mathbf{w_0 w_a w_b}$CDM model with Planck18 TTTEEE+lensing+DESI+DES-SNYR5 datasets.}
    Posterior distributions for $(w_0, w_a, w_b)$ assuming uniform priors on the  parameters.
    }
    \label{fig:quadratic}    \end{center}
\end{figure}

The discussion of the CC motivates to consider the CPL parametrization as an expansion for $w(a)$ close to $a=1$.  
In this more general framework nothing prevents the equation of state of the new species to retain values smaller than $-1$ at sufficiently early times, see \cite{Cortes:2024lgw,Shlivko:2024llw,Wolf:2023uno} for related philosophy. Within this approach, then, the CPL formula is just the linear expansion around $a=1$ of the true $w(a)$. In general,
\be\label{eq:CPLeff}
w(a)=w_0 + \sum_{n} c_n (1-a)^n = w_0 + w_a (1-a) + \frac{w_b}{2}(1-a)^2 + \cdots
\ee

Only if all the coefficients of the non-linear terms were constrained to be very small one could conclude that the CPL parametrization gives the correct evolution of $w(z)$ at all redshifts.
As we show the data  do not constrain much these terms. Many models will fall into this category and we will discuss them in Section 4.

\begin{figure}[t]
  \centering
  \includegraphics[width=0.65\textwidth]{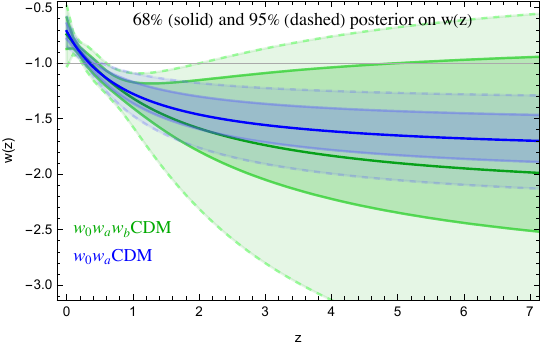}~
      \caption{\it~\textbf{Equation of state.} Marginalized constraints on the dark energy equation of state $w(z)$. We show in blue the $1\sigma$ and $2\sigma$ regions for the CPL model and  in green the extensions with quadratic coefficient $w_b$.}
  \label{fig:wofa}
\end{figure}

Truncating the series at second order the condition that $w>-1$ corresponds to 
\be\label{eq:cuts}
\begin{split}
w_b > \frac {w_a^2}{2 (w_0+1)}\,,\quad\quad w_0>-1&\quad\quad \mathrm{if}\quad w_b>0\, , \\ 
w_0+w_a + w_b/2 > -1\,,\quad\quad w_0>-1&\quad\quad \mathrm{if}\quad w_b\leq 0 \, .
\end{split}
\ee
In order to study quantitatively this scenario we have modified \texttt{CLASS} \cite{CLASS-II} to allow for this more general parametrization and sampled the posterior distributions with \texttt{Montepython} \cite{Audren:2012wb,Brinckmann:2018cvx}.  The result for the three parameters $(w_0,w_a,w_b)$, sampled without priors, are shown in figure \ref{fig:quadratic}. As anticipated the second order coefficient $w_b$ is only weakly constrained and can be positive or negative. The fit moreover does not significantly improve with the addition of the new parameter implying that the observables are not very sensitive to this extension. In Fig. \ref{fig:wofa} we derive the constraints on the effective equation of state as a function of redshift (obtained  for discrete values of $z$ from the MCMC as derived parameters).

\section{Consistent Models}\label{sec:models}

Having established the allowed regions of parameter space we now turn to models, 
 aiming at identifying non-pathological realizations based on consistent effective field theories.

The equation of state of the background can be obtained with a rolling real scalar field.\footnote{Any background with $w>-1$ could also be reproduced with 
with a perfect fluid. In this case however the adiabatic speed of sound is determined  by the equation of state as $c_s^2= w- \frac{a dw/da}{3(1+w)}$. For evolving dark energy as hinted by DESI and supernovae  the speed of sound is large and negative leading to clustering of dark energy \cite{Batista:2021uhb}. We do not consider further this possibility.}
This approach faces two main complications. As we have seen the best fits mildly prefer an equation of state where $w<-1$ at early times. 
This can be realized in k-essence type models introducing higher derivative terms. Alternatively one might include a prior  $w>- 1$
in which case the background can be realized within quintessence theories with 2-derivative effective actions.

An important phenomenological constraint arises from clustering of dark energy, see \cite{Batista:2021uhb} for a review. When $w \ne -1$ dark energy is necessarily dynamical and can thus form structures on its own. This is of course very strongly constrained by data that are instead roughly in agreement with cold DM and non-dynamical dark energy. Clustering of dark energy depends on its speed of sound. If the $c_s^2$ is large (but positive) the dark energy perturbations (read out from \eqref{perfect}) follow the gravitational potentials \cite{Batista:2021uhb},
\begin{equation}
\delta\approx -\frac {1+w}{c_s^2} \Phi\,.
\end{equation}
This leads to negligible perturbations compared to matter, that are instead proportional to $\delta_m \sim k^2 \Phi$. This is the case of quintessence models where 
the two derivative kinetic term implies $c_s^2=1$.
Alternatively if the speed of sound could be made negligible, the relevance of dark-energy clustering would need to be assessed precisely, since in this case the perturbations behave as a pressureless fluid. 

Next we address $k$-essence and $q$-essence models, with a main focus on the second.

\subsection{k-essence models}
\label{sec:k-essence}
We here outline a possible strategy to render the best-fit of CPL consistent theoretically.
A possibility is to consider scalars with higher derivative terms. 
Specifically we consider $k-$essence models \cite{Armendariz-Picon:1999hyi,Armendariz-Picon:2000nqq} described by the following effective largrangian for the real degree of freedom $\phi$
\begin{equation}
\mathscr{L}_{\rm k-essence}= P(\phi, X)\,,\quad\quad X \equiv g^{\mu\nu}\partial_\mu \phi \partial_\nu \phi\,.
\end{equation}
The novelty compared to $q-$essence models is that the general dependence on $\phi$ and $X$ allows to fit $w$ and $c_s^2$ independently. Notice that $P$ is a generic function.
Expanding around a background that is solution of the equations of motion one finds
\begin{equation}
\rho= 2 X  P_X-P\,,~~~~~~p= P \,,~~~~~~~~c_s^2= \frac {P_X}{P_X+2 P_{XX} X}\,,
\end{equation}
where $P_X\equiv \partial P/\partial X$ and $P_{XX}\equiv \partial^2 P/\partial X^2$ are derivatives of $P$ evaluated on the background solution.

The construction of a model that fits a given background goes as follows, this derivation strictly follows the discussion in \cite{Creminelli:2008wc}, see also \cite{Armendariz-Picon:1999hyi,Armendariz-Picon:2000nqq}. The relevant equation of state is reproduced by the lagrangian,
\begin{equation}
P(\phi,X)=\frac 1 2 (p-\rho)(a(\phi))+\frac 1 2 (p+\rho)(a(\phi))X+ \frac 1 2 M^4(\phi) (X-1)^2
\end{equation} 
One can check that $p$ and $\rho$ are reproduced on the background solution of the equations of motion $\phi=t$, where $t$ is the physical time.
Note that in the case of k-essence the lagrangian is not uniquely determined by the equation of state and speed of sound \cite{Unnikrishnan:2008ki}.

\begin{figure}[t]
    \centering
 

     \begin{minipage}{0.44\textwidth}
        \centering
        \includegraphics[width=\linewidth]{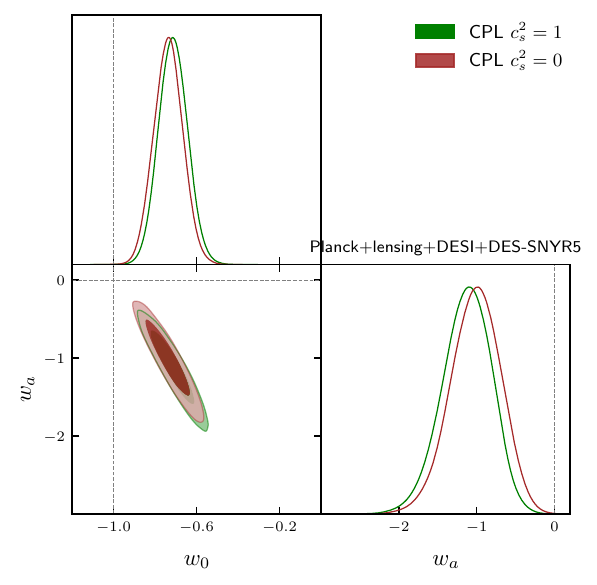}
    \end{minipage}
 
    \caption{\it~ \textbf{ Marginalized posterior distributions for $(w_0,w_a)$ of CPL model.}  We compare the case $c_s^2=0$ with $c_s^2=1$ using the prescription~\cite{Fang:2008sn} for the perturbations.
        \label{fig:cs0} }
\end{figure}

Within this approach it is possible to construct  theories that cross the phantom divide $w<-1$  without ghost particles.
The speed of sounds is given by,
\begin{equation}
c_s^2= \frac {\rho+p}{\rho+p +4 M^4}
\end{equation}
The speed of sound vanishes for $w=-1$ and becomes negative for $w< -1$ signalling a Jeans instability. 
By tuning $M$ to be hierarchically large, $c_s^2$ can be  made sufficiently small and possibly stabilized by higher derivative 
terms. To fully assess the viability of this scenario would require further study given that the clustering of dark energy would look similar to the DM one; we postpone to future work. As a preliminary investigation we have run \texttt{CLASS} with CPL scenario with $c_s^2=0$. In this case the prescription~\cite{Fang:2008sn} to cross the phantom divide is not needed and we have checked that the result actually does not depend on it. While on the best fit parameters the $\chi^2$  changes by about $4$ units compared to the standard CPL case with $c_s^2=1$ (implemented with the~\cite{Fang:2008sn} prescription),  we have run a full MCMC fit to the data, getting almost the same $\chi^2$, see the table in section~\ref{sec:conclusions} and a small shift in the parameters, see Fig.~\ref{fig:cs0}. This is a promising indication that 
k-essence scenarios can be phenomenologically viable.


\subsection{Quintessential models for DESI anomaly}

Adopting the point of view that the CPL parametrization has to be interpreted as an effective field theory expansion, or in other words as  the first term of a Taylor expansion of $w(a)$ around the present epoch, it is not difficult to identify healthy models that fit the data. The main strategy here is to consider models that never cross the phantom divide at $w=-1$. In an effective approach, they correspond to models that contribute to higher terms in the scale factor in eq.~\eqref{eq:CPLeff}.

The simplest realization are quintessence type models (see \cite{Tsujikawa:2013fta} for a review). 
They are simply defined by the following action for a real scalar field $\phi$,
\begin{equation}\label{eq:Qessence}
\mathscr{L}_{\rm q-essence}= \frac 1 2 g^{\mu\nu}\partial_\mu \phi \partial_\nu \phi - V(\phi) \,.
\end{equation}
At the level of background,
\begin{equation}
\rho =\frac {\dot{\phi}^2}2+V\,,\quad\quad p =\frac {\dot{\phi}^2}2-V\,.
\end{equation}
These models are special in two ways: first they have a built-in equation of state with $ w\geq-1$, second the sound speed is $c_s^2= 1$. 
The latter property follows from the fact that the action has just 2 derivatives.
The relativistic speed of sound implies that the perturbations do not grow inside the horizon and for this reason we do not consider the phenomenological impact of clustering of dark energy in this context.
 
In this section we would like to identify what is the function $V(\phi)$ that best fit the data. 
Interestingly, for a sensible equation of state as the one emerging from eq.~\eqref{eq:Qessence}, which is bounded by $-1\leq w \leq 1$, we can reverse engineer $w$ to extract $V$. 
As shown in Ref.~\cite{Padmanabhan:2002vv,Guo:2005ata,Barboza:2011py,Scherrer:2015tra} we can map any equation of state into a potential $V(\phi)$. The procedure only assumes the existence of a background homogenous solution $\phi(a)$ that produces the observed equation of state $w(a)$.  In this limit the potential is $V=\frac{\rho(a)}{2}\times(1-w(a))$. Therefore expressing the scale factor as
a function of $\phi$ we can reconstruct the potential. This last step is possible assuming that the energy is conserved modulo the expansion of the universe, using eq.~\eqref{eq:f}. This implies that the quintessential sector has to be isolated and in particular not exchange energy with DM. With these assumptions one can write
\be
\frac{d\phi}{da}=\frac{\sqrt{(1+w(a))\rho(a)}}{a H(a)}\,.
\ee
By integrating the above expression and solving for $a(\phi)$, the potential is determined as function of $\phi$. 
In formulae, making connection with our notation, we get the following numerical prescription
\begin{equation}
\left\{\begin{split}
&\frac{V(\phi)}{\rho_\Lambda}= \frac12  (1- w_{\rm DE}(a(\phi)) f_{\rm DE}(a(\phi))\,,  \\
&\frac{\phi(a)}{\Mpl}=-\int_a^1 \frac{\sqrt{3(1+w_{\rm DE}(\tilde a))\Omega_{\rm DE} f_{\rm DE} (\tilde a)}}{\sqrt{\Omega_{m}/\tilde a^3+  \Omega_{\rm DE} f_{\rm DE}(\tilde a)}} \frac {d\tilde a}{\tilde a} \,,
\end{split}\right.
\label{eq:metodo}
\end{equation}
where $\rho_\Lambda$ is the quintessence energy density today and we have approximated $H(a)$ to the expression where the only energy components at late times are matter and dark energy. 
$f_{\rm DE}$ can be derived solving the continuity equation that is equivalent to the equation of motion for the rolling scalar field. Note that the integral in the second line converges for $a\to 0$ so that the field range for $\phi$ is finite, typically a fraction of the (reduced) Planck mass $M_{\rm Pl}$. In terms of initial conditions this implies that $V(\phi)$ gives the correct equation of state if $\phi$ starts from $\phi_0\equiv \phi(a= 0)$ with null velocity. This procedure does not determine the field for $\phi<\phi_0$. In practice for a generic equation of state, $V(\phi)$ can be easily determined numerically.


\paragraph{The Ramp}~\\

\begin{figure}[t!]
\begin{center}
\begin{tikzpicture}[line width=1.2 pt, scale=1.7]

\fill [shade,color=lightgray] (-1,0) rectangle (3.6,-0.4);

  \draw[stealth-, color=black] (-1,0) -- (3.6, 0);
  \draw[-stealth, color=black] (0,-0.4) -- (0, 2.5);
   

 \draw[color=darkgreen] (0,.3) -- (2.5, .3);
 \draw[color=darkgreen] (2.5, .3) -- (3, 2);

 \draw[line width=0.7pt, dashed, color=gray] (0,2) -- (3,2);
 \draw[line width=0.7pt, dashed, color=gray] (2.5,.3) -- (2.5,0);
 \draw[line width=0.7pt, dashed, color=gray] (3,2) -- (3,0);
  
  	\node at (-0.5,0.2) {\small redshift $z$};
	\node at (1.3,-0.2) {$w<-1$};
	\node at (1.5,2.2) {\textcolor{darkgreen}{$w(z)$}};
	\node at (-0.2,.425) {$w_i$};
	\node at (-0.2,2) {$w_0$};
	\node at (2.35, 0.1) {\small $z_s$};
	\node at (3.3,0.12) {\small today};

 \end{tikzpicture}
    \includegraphics[width=0.4\linewidth]{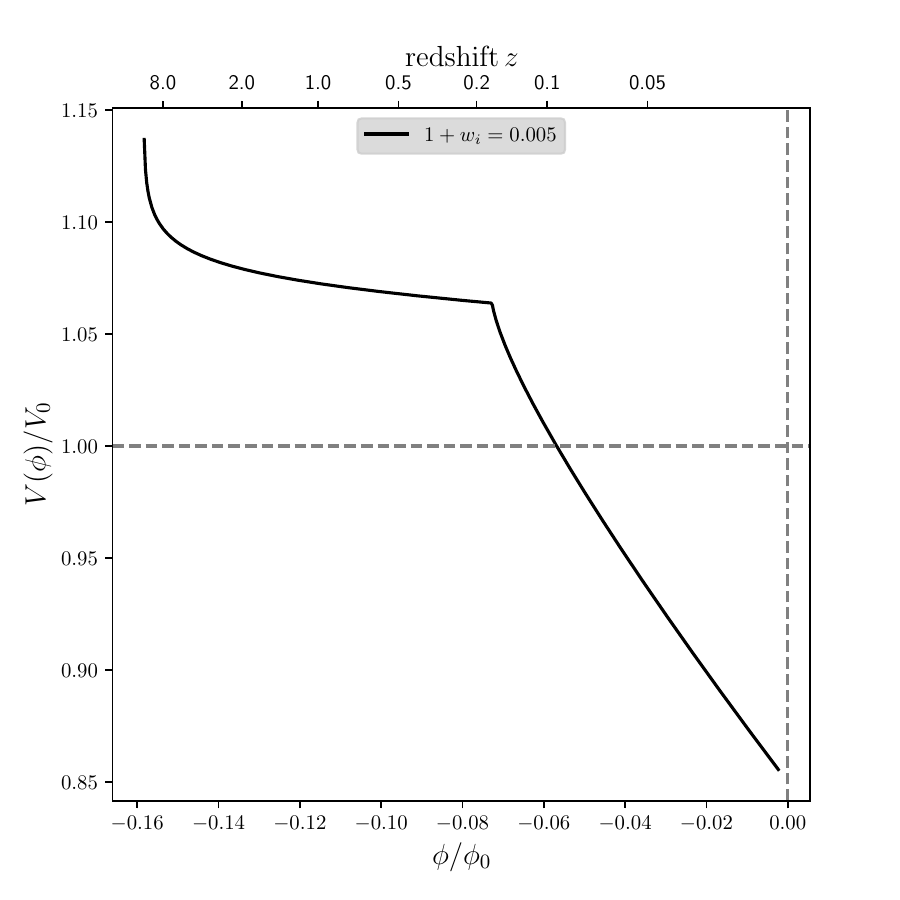}
    \caption{\it~  \label{fig:ramp} Left: Equation of state of the Ramp model. Right: the quintessence potential reproducing the best-fit Ramp model, as a function of redshift. Note that the shape at high redshift is not constrained by data and it depends on the initial equation of state, which has been fixed here to $1+w_i=0.005$.} 
\label{fig:ramp}

\end{center}
\end{figure}

As a simple example we consider a CPL-like model that never crosses $w=-1$.
To do this we simply assume that at early times the equation of state becomes constant, $w_i>-1$,
\footnote{Strictly speaking an exact equation $w_i=-1$  is not consistent as a dynamical model because no degrees of freedom propagate in this limit. 
We find however that $w_i=-1$ is an excellent approximation to compute relevant observables.}
\be\label{eq:rampa}
w_{\rm RAMP}=\mathrm{max}[w_i, w_0 +w_a(1-a)]\,.
\ee
The scale factor where the equation of state changes behavior is given by,
\be
a_s\equiv \frac 1 {1+z_s}=\frac{w_0+w_a-w_i}{w_a}\,.
\ee
By solving eq.~\eqref{eq:f} one finds that the energy density evolves as,
\begin{equation}
f_{\rm ramp}(a)= f_{\rm CPL}(a) \theta(a-a_s)+ f_{\rm CPL}(a_s)\left(\frac {a_s}{a}\right)^{3(1+w_i)}\theta(a_s-a) \, ,
\end{equation}
where $f_{\rm CPL}(a)$ is found in eq.~\eqref{eq:fCPL} and $\theta$ here is a step function. The evolution of $w_{\rm RAMP}(a)$ is shown in figure \ref{fig:ramp}.

We implemented this model in \texttt{CLASS}, in order to perform a fit to the data.
Since $w\geq -1$ we can consistently include perturbations of the scalar field~\footnote{In practice we have used $w_i=-1+\epsilon$ in the code, and checked that results do not depend on $\epsilon$, as long as $0<\epsilon \ll 0.1$.}. 
However for $w_i\approx -1$ these are very suppressed for the modes the cross the horizon for $a< a_s$.

As a consequence the result is not sensitive to the presence of perturbations; to show that this is the case we have checked that with a different speed of sound, i.e. $c_s=0$, there is a negligible impact on the $\chi^2$ for the Ramp model.  This should be compared with the CPL parametrization where instead changing to $c_s=0$  induces a non-negligible shift in the fit, as discussed in section 4.1, of $\Delta\chi^2\approx 4$ on the best-fit model, mostly due to the low-$\ell$ Temperature spectra.

\begin{figure}[p]
\begin{center}
\begin{minipage}{.9\textwidth}
\centering
\includegraphics[width=.8\textwidth]{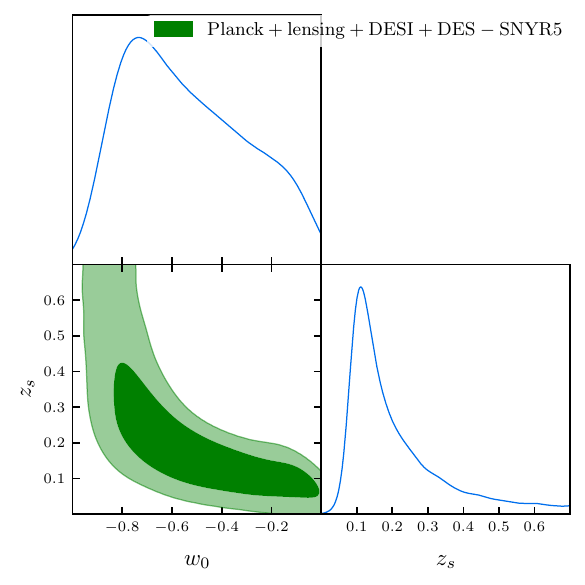}~
\end{minipage}

\begin{minipage}{\textwidth}
    \centering
    \renewcommand{\arraystretch}{1.1} 
    \begin{tabular}{l|c|c|c|c}
        \hline
        \textbf{Parameter} & \textbf{Best-fit} & \textbf{Mean $\pm \sigma$} & \textbf{95\% lower} & \textbf{95\% upper} \\ \hline
       
$100~\omega_{\rm b }$ &$2.256$ & $2.249_{-0.014}^{+0.014}$ & $2.222$ & $2.276$ \\ 
$\omega_{\rm cdm }$ &$0.1183$ & $0.1181_{-0.00091}^{+0.00092}$ & $0.1163$ & $0.1199$ \\ 
$100*\theta{}_{s }$ &$1.042$ & $1.042_{-0.00029}^{+0.00029}$ & $1.041$ & $1.043$ \\ 
$\log(10^{10}A_{s })$ &$3.059$ & $3.054_{-0.016}^{+0.014}$ & $3.024$ & $3.085$ \\ 
$n_{s }$ &$0.9709$ & $0.9688_{-0.0038}^{+0.0038}$ & $0.9614$ & $0.9762$ \\ 
$\tau{}_{reio }$ &$0.06355$ & $0.06093_{-0.0082}^{+0.0071}$ & $0.04596$ & $0.07645$ \\ 
\hline
$w_0$ &$-0.65$ & $-0.5387_{-0.36}^{+0.16}$ & $-0.9196$ & $-0.04705$ \\ 
$z_s$ &$0.1967$ & $0.2521_{-0.21}^{+0.031}$ & -- & -- \\ 
\hline\hline
$\Omega_{\rm CPL}$ &$0.6784$ & $0.6769_{-0.007}^{+0.0073}$ & $0.6629$ & $0.6907$ \\ 
$H_0$[km/s/Mpc] &$66.34$ & $66.15_{-0.65}^{+0.63}$ & $64.9$ & $67.4$ \\ 
$M$[DES nuisance] &$-0.07516$ & $-0.07753_{-0.011}^{+0.011}$ & $-0.0987$ & $-0.05645$ \\ 
 \hline\hline
                \multicolumn{5}{c}{$\llike = 2218$} \\ \hline
    \end{tabular}
\end{minipage}
       \caption{\it~ \textbf{The Ramp model with Planck18 TTTEEE+lensing+DESI+DES-SNYR5 datasets, with no perturbations in the dark energy fluid.}
    Marginalized posterior distributions for $(w_0,z_s)$.  We assume priors $w_0\in [-1,0]$ and $z_s\in[0,\infty)$, and uniform priors for other parameters. }
    \label{fig:rampa}
   \end{center}
\end{figure}

We perform a fit to the full dataset,  by scanning over $w_0$ and $z_s$, since this guarantees a better convergence of the MCMC chains. The results of the numerical analysis are found in figure \ref{fig:rampa}. We notice that $\llike$ is slightly larger than the CPL model ($w_0w_a$CDM),  The best fit values are $(w_0,z_s)_{\rm best}=(-0.65,0.20)$, signalling that the transition from the CC is rather abrupt at $z_s$, since in terms of $w_a$ this corresponds to $w_a\approx -2.1 $. This same parameter point for $w_0,w_a$ is marginally consistent with data when used in the CPL model  (see fig.~\ref{fig:validation}).

\begin{figure}[t]
  \centering
  \includegraphics[width=0.8\textwidth]{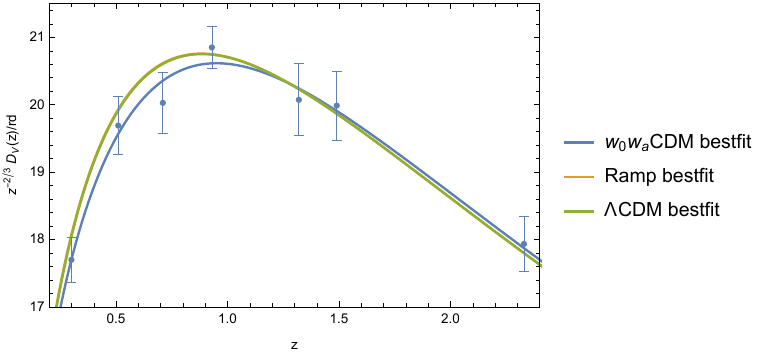}~
      \caption{\it~\textbf{Best fit of ramp model.} Projection of the best fit on the DESI data. We plot the angle averaged distance $D_V(z)=(z D_M(z)^2 D_H(z))^{1/3}$.
      For comparison the best fit of CPL  and $\Lambda$CDM models are also shown.}
  \label{fig:bestfit}
\end{figure}

We conclude that the ramp scenario provides an excellent fit of the current data in a consistent quantum field theory framework. 
The minimum $\chi^2$ is slightly worse than the CPL model but the latter is pathological in single field models and it is adjusted by ad-hoc perturbations that are coming from multiple field scenarios, to cross the phantom divide. In Fig. \ref{fig:bestfit} we compare the best fit of the ramp model with DESI data. Note that the DESI fit is very close to the $\Lambda$CDM, and indeed we have checked that the smaller $\chi^2$ is mostly due to the Supernova dataset.

Let us comment that while this work was in preparation, other studies of quintessence in this context were considered. In particular in \cite{Ramadan:2024kmn,Bhattacharya:2024hep,Andriot:2024jsh,Tada:2024znt} the exponential quintessence was compared with data. Such fits have a worse fit than the CPL model, not surprisingly given that their equations of state differs from CPL at late times. 
Similarly the DESI collaboration~\cite{DESI:2024kob} has analyzed some one-parameter models with healthy equations of state, such as the so-called thawing model, which improves slightly over the $\Lambda$CDM fit, with $\Delta\chi^2=-5$, as opposed to  $\Delta\chi^2=-18$ for CPL. More general equations of state were 
also considered by DESI collaboration in \cite{DESI:2024aqx}. More recently in~\cite{Gialamas:2024lyw} quintessence models with a step in the equation of state and a fit comparable to CPL was also considered. 

Finally we also note that none of the late-time dark energy models studied here improve significantly on the Hubble tension with local measurements by the SH$0$ES collaboration~\cite{Riess:2021jrx}, when including DESI+ Planck18+ Supernova datasets (Pantheon$+$ or DES-SNYR5). This is true also for the $w_0w_a$CDM model, see~\cite{DESI:2024mwx}. In order to find models that address the $H_0$ tension one may instead modify the early universe and the sound horizon $r_d$: in recent work it has indeed been shown that the addition of a Dark radiation fluid significantly reduces the tension~\cite{Allali:2024cji}.

\section{Conclusions}
\label{sec:conclusions}

By far the simplest explanation for the acceleration of the universe is the cosmological  constant, even if extremely fine-tuned. 
Any time-dependence of this number would shake our understanding of the universe. 
It is thus very interesting that recent experiments hint to a more complex dark energy than $\Lambda$. 
In particular recent DESI BAO + Planck18 CMB data + Supernovae measurements appear to favour 
evolving dark energy compared to cosmological constant.

In this work we have broadly studied possible interpretations of data-driven dynamical dark energy models, within theoretically consistent extensions of $\Lambda$CDM. 
Our first observation is that for the best fit the CPL equation of state crosses the phantom divide $w=-1$ where theoretical
consistency of the theory is far from obvious and thus this motivates us to consider more general parametrizations where the equation of
state deviates from the CPL parametrization at early times. This is also motivated by an effective field theory perspective where 
the CPL equation of state describes only the first term of a Taylor expansion in the scale factor around the present epoch.
Not surprisingly higher order terms are weakly constrained by current observations that mostly test the universe at late times, i.e. small redshift. We perform in particular the cosmological fit including a quadratic term in the equation of state. Given that the data mostly constrain
the dark energy at low redshift this is expected to capture many models beyond the CPL fluid.

The addition of higher order terms opens the way to the construction of models that provide a fit of DESI data comparable to 
the CPL parametrization while avoiding  $w<-1$. In particular we show that this can be realized in quintessence 
theories, also constructing a potential that allows for a good fit the data. In the simplest case studied in this paper -- the ramp model -- we study the  
quintessence model closest to CPL where  the equation 
of state coincides with CPL at late times and approaches $w\approx -1$ at early times. This scenario features the same number parameters of the CPL parametrisation
and still has a large improvement on the $\llike$ with respect to $\Lambda$CDM, only slightly worse than $w_0w_a$CDM, but with the crucial advantage of being described by a simple and theoretically consistent lagrangian. 

Our survey of models is summarized by the following table: 
\begin{center}
\begin{tabular}{>{\columncolor{gray!30}}>{\centering\arraybackslash}p{3cm}||>{\centering\arraybackslash}p{2.5cm}|>{\centering\arraybackslash}p{2.5cm}|>{\centering\arraybackslash}p{2.5cm}||>{\centering\arraybackslash}p{2cm}||>{\centering\arraybackslash}p{2cm}}
\hline
\rowcolor{green!20}
\textbf{$w_0w_a$CDM} & $w_0$ & $w_a$ & -- & $H_0$\scriptsize[km/s/Mpc] & $\Delta \chi^2$ \\
\hline
\rowcolor{red!10}
$c_s^2=1$ & $-0.7124_{-0.073}^{+0.069}$  & $-1.13_{-0.29}^{+0.35}$ & --- & $67.43_{-0.67}^{+0.65}$  & $-18$ \\
$c_s^2=0$ &  $-0.7354_{-0.072}^{+0.07}$   & $-1.018_{-0.31}^{+0.34}$ & --- & $67.31_{-0.69}^{+0.65}$   & $-17$ \\
\hline
\end{tabular}
\begin{tabular}{>{\columncolor{gray!30}}>{\centering\arraybackslash}p{3cm}||>{\centering\arraybackslash}p{2.5cm}|>{\centering\arraybackslash}p{2.5cm}|>{\centering\arraybackslash}p{2.5cm}||>{\centering\arraybackslash}p{2cm}||>{\centering\arraybackslash}p{2cm}}
\hline
\rowcolor{green!20}
\textbf{$w_0w_a\Lambda$CDM} & $w_0$ & $w_a$ & $\Omega_\Lambda$ & $H_0$\scriptsize[km/s/Mpc] & $\Delta \chi^2$ \\
\hline
$c_s^2=1$ & $-0.7425_{-0.14}^{+0.16}$ & $-2.359_{-0.61}^{+1.8}$ &  $0.237_{-0.054}^{+0.26}$ &$67.32_{-0.65}^{+0.62}$ & $-18$ \\
\hline
\end{tabular}
\begin{tabular}{>{\columncolor{gray!30}}>{\centering\arraybackslash}p{3cm}||>{\centering\arraybackslash}p{2.5cm}|>{\centering\arraybackslash}p{2.5cm}|>{\centering\arraybackslash}p{2.5cm}||>{\centering\arraybackslash}p{2cm}||>{\centering\arraybackslash}p{2cm}}
\hline
\rowcolor{green!20}
\textbf{$w_0w_a w_b$CDM} & $w_0$ & $w_a$ & $w_b$ & $H_0$\scriptsize[km/s/Mpc] & $\Delta \chi^2$ \\
\hline
$c_s^2=1$  &  $-0.7425_{-0.14}^{+0.16}$ & $-0.7822_{-1.6}^{+1.2}$  & $-1.465_{-3.7}^{+5.8}$ & $67.42_{-0.76}^{+0.78}$   & $-18$ \\
\hline
\end{tabular}
\begin{tabular}{>{\columncolor{gray!30}}>{\centering\arraybackslash}p{3cm}||>{\centering\arraybackslash}p{2.5cm}|>{\centering\arraybackslash}p{2.5cm}|>{\centering\arraybackslash}p{2.5cm}||>{\centering\arraybackslash}p{2cm}||>{\centering\arraybackslash}p{2cm}}
\hline
\rowcolor{green!20}
\textbf{Q-ess. ramp} & $w_0$ & $z_s$ & -- & $H_0$\scriptsize[km/s/Mpc] & $\Delta \chi^2$ \\
\hline
$c_s^2=1$  & $-0.5387_{-0.36}^{+0.16}$ &  $0.2521_{-0.21}^{+0.031}$ & --- & $66.15_{-0.65}^{+0.63}$ & $-12$ \\
\hline
\end{tabular}\\
\end{center}
where we report $\Delta \chi^2\equiv \chi^2_{\rm model}-\chi^2_{\Lambda {\rm CDM}}$ and  $\chi^2 = -2\log\mathscr{L}_{\rm min}$.

All the models above perform significantly better than $\Lambda$CDM from the point of view of data, but only the Quintessence ramp model can be considered as a consistent theory.
Future data from DESI, Euclid~\cite{Euclid:2024yrr}  and supernovae will tell if evolving dark energy is fact or fiction.

\subsubsection*{Acknowledgements}
We acknowledge the use of the computing resources provided by the ``PC-Farm'' at INFN Florence. 
The plots of the marginalized posterior distributions in all the figures are done with \texttt{GetDist} \cite{getdist}.
We wish to thank Marko Simonovic for discussions on dark energy and large scale structure.

\appendix

\section{Validation of DESI results}\label{appA}
In this appendix we reproduce the DESI analysis to validate our study of more general variations of $\Lambda$CDM. The workflow of our analyses is the following. First we implement in \texttt{CLASS} \cite{CLASS-II} the modified cosmology of interest with appropriate cosmological parameters, then we perform a Monte-Carlo Markov-Chain (MCMC) bayesian inference of posterior distributions of such parameters by using the tool \texttt{MontePython} \cite{Audren:2012wb,Brinckmann:2018cvx}. We use data and likelihood as follows
\begin{itemize}
\item \textit{Planck18}. We use the full dataset of Planck 2018 \cite{Planck:2019nip}, in high-$\ell$ and low-$\ell$ for $TT, TE, EE$ and lensing. We have sampled all the nuisance parameters of the likelihoods.
\item \textit{DESI}.  For DESI-Y1  we used the values of $D_{M}, D_H$ and $D_V$ reported in Table 1 of \cite{DESI:2024mwx}. We constructed the covariance matrix following the correlations provided in that table.
\item \textit{Pantheon+}. We use the SN catalogue of \cite{Pantheon+}.
\item \textit{DES-SNYR5}. We use the SN dataset of \cite{DES:2024tys}. We downloaded the data and the covariance matrix from \cite{DES-data} and we adapted the likelihood  code to \texttt{MontePython}. We have checked that we are able to reproduce the results of \cite{DES:2024tys}, more details can be found in the appendix \ref{app:inputs} and figure \ref{fig:check_DES-SNYR5}.
\end{itemize}
We adopt the standard choice of scanning over the following cosmological parameters (with $\Omega_k=0$)
\be
\{100~\omega_{\rm b},\quad  \omega_{cdm},\quad 100\,\theta_s, \quad \log(10^{10}A_{s}), \quad n_s,\quad \tau_{\rm reio}\},
\ee
plus any other extra parameter from the model under consideration (see Appendix \ref{app:inputs} for more details). In particular for the CPL model under consideration in this section we have added $w_0$ and $w_a$ to the cosmological parameter list. 

\begin{figure}[p]
\begin{center}
\begin{minipage}{.9\textwidth}
\centering
\includegraphics[width=.8\textwidth]{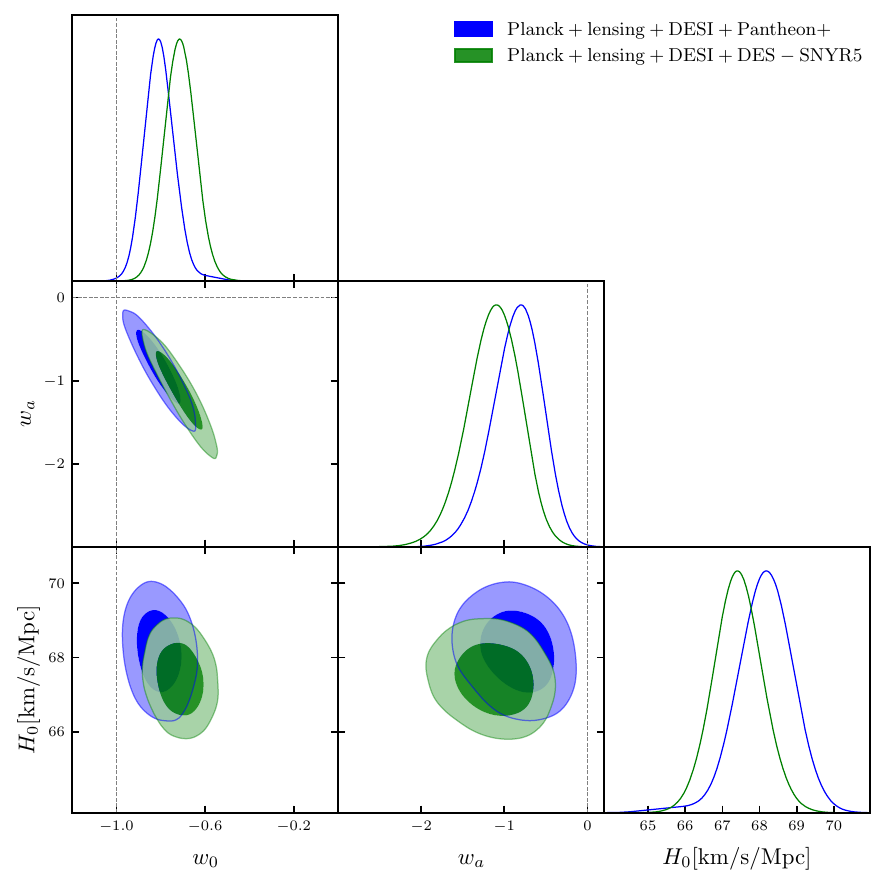}~
\end{minipage}

\begin{minipage}{\textwidth}
    \centering
    \renewcommand{\arraystretch}{1.1} 
    \begin{tabular}{l|c|c|c|c}
        \hline
        \textbf{Parameter} & \textbf{Best-fit} & \textbf{Mean $\pm \sigma$} & \textbf{95\% lower} & \textbf{95\% upper} \\ \hline
        $100~\omega_{b}$ & $2.245$ & $2.239_{-0.015}^{+0.014}$ & $2.211$ & $2.266$ \\ 
        $\omega_{\rm cdm}$ & $0.1199$ & $0.1198_{-0.001}^{+0.001}$ & $0.1178$ & $0.1218$ \\ 
        $100~\theta_{s}$ & $1.042$ & $1.042_{-0.00028}^{+0.00029}$ & $1.041$ & $1.042$ \\ 
        $\ln(10^{10}A_{s})$ & $3.031$ & $3.043_{-0.015}^{+0.014}$ & $3.014$ & $3.072$ \\ 
        $n_{s}$ & $0.9657$ & $0.9658_{-0.0039}^{+0.0039}$ & $0.9582$ & $0.9733$ \\ 
        $\tau_{\rm reio}$ & $0.04764$ & $0.05395_{-0.0075}^{+0.0072}$ & $0.0392$ & $0.06904$ \\ 
     \hline
        $w_{0}$ & $-0.6883$ & $-0.7124_{-0.073}^{+0.069}$ & $-0.8505$ & $-0.5733$ \\ 
        $w_{a}$ & $-1.255$ & $-1.13_{-0.29}^{+0.35}$ & $-1.774$ & $-0.508$ \\ 
        \hline\hline
        $\Omega_{\rm CPL}$ & $0.687$ & $0.6858_{-0.0066}^{+0.0066}$ & $0.6729$ & $0.6989$ \\ 
        $H_0$[km/s/Mpc] & $67.6$ & $67.43_{-0.67}^{+0.65}$ & $66.13$ & $68.75$ \\ 
        $M$ [DES nuisance] & $-0.03325$ & $-0.04001_{-0.017}^{+0.017}$ & $-0.0733$ & $-0.006517$ \\ \hline\hline
                \multicolumn{5}{c}{$\llike = 2215$} \\ \hline
    \end{tabular}
\end{minipage}
 \caption{\it~  \label{fig:validation} \textbf{$\mathbf{w_0 w_a}$CDM model with Planck18 TTTEEE+lensing+DESI+DES-SNYR5 and Pantheon+ datasets. }
    Marginalised posterior distributions for $w_0$ and $w_a$ (blue, Pantheon+, green DES-SNYR5).  
    As in the rest of this work we use for simplicity fixed neutrino masses to the minimum value from neutrino oscillation measurements, $\sum m_\nu=0.06$ eV,  and  we assume uniform prior on the input parameters.
The table refers to the best fit using only DES-SNYR5.  }
    \end{center}
\end{figure}

We show the marginalized posterior distribution of the CPL model in Fig. \ref{fig:validation} that fully agrees with~\cite{DESI:2024mwx}.
The minimum of the (minus) logarithm of the likelihood, $\llike$, using the DES-SNYR5 supernovae dataset is found to be 2215 while for $\Lambda$CDM the value would be 2224. For a Gaussian distribution this corresponds to a $\sim 4 \sigma$ deviation. We note however that the CPL fit benefits from the inclusion of perturbations according to the prescription of \cite{Fang:2008sn} which is actually based on a multifield setup, in order to cross the phantom divide.

\subsection{Results for CPL parametrization}
In this section we summarize our results for the CPL model, also referred to as $w_0w_a$CDM model in the literature. The posterior distributions for $(w_0, w_a)$ marginalized over the remaining parameters are shown in figure \ref{fig:validation}, together with best fit  values and $95\%$ CL intervals of all the relevant parameters.

They can be compared with the results from \cite{DESI:2024mwx}, finding agreement. In particular we have found $\Lambda$CDM to be disfavoured at the same level. We take this as a guideline for future searches in this paper.

\section{Inputs for the MCMC}

In this appendix we collect some technical material regarding: $1)$ the use of the DES-SNYR5 datasets; $2)$ a comment on the approach to CMB data without the use of the full {\it Planck18} likelihood.

\label{app:inputs}

\subsection{Supernovae datasets}
We have used both \textit{Pantheon+} and \textit{DES-SNYR5} as catalogues of supernovae. The likelihood of \textit{Pantheon+} is available in \texttt{MontePython}, while we have constructed the likelihood for \textit{DES-SNYR5}, as was done also in~\cite{Allali:2024cji}, starting from the dataset available in~\cite{DES-data} and following the \texttt{Python} script available in  \texttt{MontePython} for \textit{Pantheon+},  based on \cite{likePP,likeShoes}. 

\begin{figure}[t]
    \centering
     \begin{minipage}{0.45\textwidth}
        \centering
        \includegraphics[width=\linewidth]{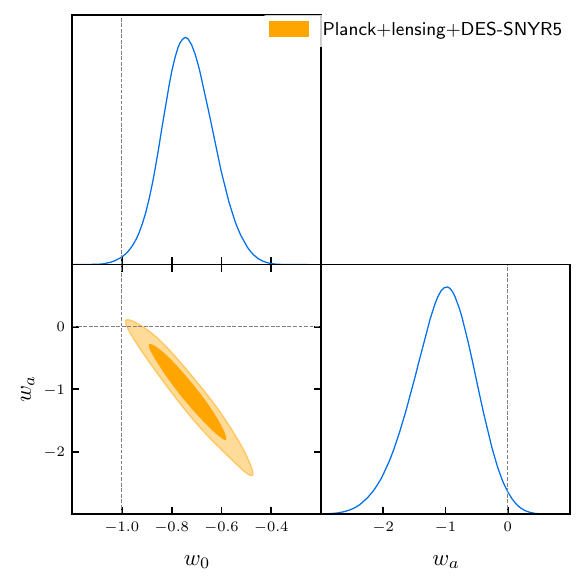}
    \end{minipage}
    \caption{\it~ \textbf{$\mathbf{w_0 w_a}$CDM model with Planck18 TTTEEE+lensing+DES-SNYR5 datasets.} Marginalized posterior distributions for $(w_0,w_a)$. This result has to be compared with figure 8  of Ref.~\cite{DES:2024tys}, with the caveats discussed in the text.  }
    \label{fig:check_DES-SNYR5}
\end{figure}

The likelihood can be used for the MCMC sampling. Here we report the computation of the posterior distribution for the CPL model done with full {\it Planck18} dataset and DES-SNYR5. The results are shown in figure \ref{fig:check_DES-SNYR5}. We notice however that we have included the full dataset, while in \cite{DES:2024tys} $Planck$-$lite$2015 was used.

\subsection{Simplified CMB+lensing likelihood}
In many cases in order to have a preliminary survey of models we have adopted a simplified approach to the CMB dataset, see for example \cite{BOSS:2014hhw} for a similar approach. This is motivated by two main reasons. The first is that we expect the fluctuations of the new component to have a small impact on the bulk of the CMB data. The second is that the tension with $\Lambda$CDM is already somewhat present in the data even without including the {\it Planck18} likelihoods (with lensing), see \cite{DESI:2024mwx}. We have then constructed a likelihood starting from the bestfit and covariance matrix of the 2018 full Planck dataset (including lensing) as available from \texttt{Montepython}.
Assuming gaussian distributions we have deduced a reduced covariance matrix $\Sigma$ for $\omega_b,\omega_{\rm cdm}, 100\theta_\star$, given by
\be\label{eq:simple-like}
\begin{split}
&\Sigma^{-1}=\left[
\begin{array}{ccc}
3.48894 \times 10^{7} & 2.13917 \times 10^{6} & -1.63347 \times 10^{6} \\
2.13917 \times 10^{6} & 496795 & 254184 \\
-1.63347 \times 10^{6} & 254184 & 6.13782 \times 10^{6}\end{array}
\right]\\
& \quad \quad [\omega_b,\omega_{\rm cdm}, 100\theta_\star]_{\rm best}=[0.0223622,0.120167,1.04171] \, ,
\end{split},
\ee
which has been used to construct a gaussian likelihood with the best-fit values above. Performing a fit to the CPL model ($w_0w_a$CDM), we get a minimum
 $-\log\mathcal{L}_{\rm min}=826$ for the datasets CMB-simple+DESI+DES-SNYR5.  However, notice that in the main text all our results are based on the full PLANCK 2018 likelihood (TTTEEE, high/low-$\ell$ + lensing) as discussed in section \ref{sec:data}.

\pagestyle{plain}
\bibliographystyle{jhep}
\small
\bibliography{biblio}

\end{document}